\documentclass{jfm}
\usepackage{graphicx,graphics}
\usepackage{natbib}
\usepackage{color}
\usepackage[hyphens]{url}
\usepackage{amssymb,amsmath,amsfonts}
\DeclareGraphicsExtensions{.jpg,.jpeg,.pdf,.png,.mps,.eps,.ps,.gif}
\usepackage{epstopdf}

\ifCUPmtlplainloaded \else
  \checkfont{eurm10}
  \iffontfound
    \IfFileExists{upmath.sty}
      {\typeout{^^JFound AMS Euler Roman fonts on the system,
                   using the 'upmath' package.^^J}%
       \usepackage{upmath}}
      {\typeout{^^JFound AMS Euler Roman fonts on the system, but you
                   dont seem to have the}%
       \typeout{'upmath' package installed. JFM.cls can take advantage
                 of these fonts,^^Jif you use 'upmath' package.^^J}%
      }
  \else
  \fi
\fi

\ifCUPmtlplainloaded \else
  \checkfont{msam10}
  \iffontfound
    \IfFileExists{amssymb.sty}
      {\typeout{^^JFound AMS Symbol fonts on the system, using the
                'amssymb' package.^^J}%
       \usepackage{amssymb}%
         \let\leq=\leqslant
         \let\geq=\geqslant
      }{}
  \fi
\fi

\ifCUPmtlplainloaded \else
  \IfFileExists{amsbsy.sty}
    {\typeout{^^JFound the 'amsbsy' package on the system, using it.^^J}%
     \usepackage{amsbsy}}
    {\providecommand\boldsymbol[1]{\mbox{\boldmath $##1$}}}
\fi

\providecommand\bnabla{\boldsymbol{\nabla}}
\providecommand\bnabla{\boldsymbol{\nabla}}

\newcommand\Rey{\mbox{\textit{Re}}}  

\newsavebox{\astrutbox}
\sbox{\astrutbox}{\rule[-5pt]{0pt}{20pt}}

\newcommand{\rem}[1]{}
\DeclareMathAlphabet{\mathbi}{OML}{cmm}{b}{it}

\newcommand{\bx}{\mathbi{x}}

\newcommand{\bel}{\begin{equation}\label}
\newcommand{\ee}{\end{equation}}
\newcommand{\beq}{\begin{eqnarray}\label}
\newcommand{\eeq}{\end{eqnarray}}
\newcommand{\bc}{\begin{center}}
\newcommand{\ec}{\end{center}}
\newcommand{\ben}{\begin{enumerate}}
\newcommand{\een}{\end{enumerate}}
\newcommand{\bit}{\begin{itemize}}
\newcommand{\eit}{\end{itemize}}
\newcommand{\I}{\int_{\mathcal{V}}}

\newcommand{\bu}{\mbox{\boldmath$u$}}
\newcommand{\bv}{\mathbi{v}}

\newcommand{\bom}{\mbox{\boldmath$\omega$}}
\newcommand{\bj}{\mbox{\boldmath$j$}}
\newcommand\shalf{\ensuremath{{\scriptstyle\frac{1}{2}}}}

\newcommand\twothirds{\ensuremath{{\scriptstyle\frac{2}{3}}}}

\newcommand{\rhos}{\rho^{*}}

\title[Nonlinear effects in buoyancy-driven variable density turbulence]
{Nonlinear effects in buoyancy-driven variable density turbulence}

\author[P. Rao, C. P. Caulfield \& J. D. Gibbon]%
{P. Rao$^1$%
  \thanks{Email address for correspondence\,: prao@ams.sunysb.edu},\ns
C. P. Caulfield$^{2,3}$ \& J. D.  Gibbon$^4$}

\affiliation{$^1$Department of Applied Mathematics and Statistics, State University of New York,
Stony Brook, NY 11790, USA\\[\affilskip]
$^2$BP Institute, University of Cambridge, Madingley Rise, Madingley
Road, Cambridge CB3 0EZ, UK\\[\affilskip]
$^3$ Department of Applied Mathematics \& Theoretical Physics, University of
Cambridge, Centre for Mathematical Sciences, Wilberforce Road, Cambridge CB3 0WA, UK \\[\affilskip]
$^4$Department of Mathematics, Imperial College London, London SW7 2AZ, UK}

\pubyear{2010}
\volume{650}
\pagerange{119--126}
\date{\today}
\begin{document}

\maketitle


\begin{abstract}
We consider the time-dependence of a hierarchy of scaled $L^{2m}$-norms $D_{m,\omega}$ and $D_{m,\theta}$ of the vorticity $\bom = 
\bnabla \times \bu$ and the density gradient $\bnabla \theta$,  where $\theta=\log (\rhos/\rhos_0)$, in a buoyancy-driven 
turbulent flow as simulated by \cite{LR2007}. $\rhos(\bx,\,t) $ is the composition density of a mixture of two incompressible miscible fluids 
with fluid densities $\rhos_2 > \rhos_1$ and $\rhos_{0}$ is a reference normalisation density. Using data from the publicly available Johns 
Hopkins Turbulence Database we present evidence that the $L^{2}$-spatial 
average of the density gradient $\bnabla \theta$ can reach extremely large values, even in flows with low Atwood number $At = (\rhos_{2} - 
\rhos_{1})/(\rhos_{2} + \rhos_{1}) = 0.05$, implying that very strong mixing of the density field at small scales can arise in buoyancy-driven 
turbulence. This large growth raises the possibility that the density gradient $\bnabla\theta$ might blow up in a finite time.
\end{abstract}

\section{Introduction}

The irreversible mixing at a molecular level of two fluids of different densities $\rhos_{2} > \rhos_{1}$ 
is a fluid dynamical process of great fundamental  interest and practical importance, especially when 
the fluids are turbulent. Such turbulent mixing flows occur in many different circumstances. A particularly 
important class arises when the buoyancy force associated with the effects of statically unstable variations 
in fluid density in a gravitational field actually drives both the turbulence and the ensuing mixing 
itself. Such flows, commonly referred to as `Rayleigh-Taylor instability' (RTI) flows due to the form of the initial 
linear instability (\cite{R1900,T1950}), have been very widely studied 
(see \cite{S1984,Youngs1984,Youngs1989,Glimm2001,D2004,D2005,Lee2008,Hyunsun2008,AD2010}),  
not least because of their relevance in astrophysics \citep{Cabot2006a} and fusion \citep{P1994}.

A key characteristic of RTI flows is that the turbulence which develops is not driven by some external forcing 
mechanism, but rather is supplied with kinetic energy by the conversion of `available' potential energy stored 
in the initial density field. This kinetic energy naturally drives turbulent disorder and a cascade to small scales, 
with an attendant increase in the dissipation rate of kinetic energy. Such small scales also lead to `filamentation', 
i.e. enhanced surface area of contact between the two miscible fluids and, crucially, substantially enhanced
gradients in the density field, which thus also leads to irreversible mixing, and hence modification in the 
density distribution. There has been an explosion in interest in investigating the `efficiency' of this mixing, i.e. 
loosely, the proportion of the converted available potential energy which leads to irreversible mixing, as opposed
to viscous dissipation, (see the recent review of \cite{T2013}), although the actual definition and calculation
of the efficiency is subtle and must be performed with care -- see for example \cite{DWD2014} for 
further discussion.

Nevertheless, there is accumulating evidence that buoyancy-driven turbulence is particularly efficient in driving 
mixing \citep{LD2011,DWD2014} and certainly more efficient than externally forced turbulent flow. This evidence 
poses the further question whether there are some distinguishing characteristics of the buoyancy-driven turbulent 
flow that are different from the flow associated with an external forcing, in particular whether these characteristics 
can be identified as being responsible for the enhanced and efficient mixing. 

The situation is further complicated by the observation that, even when the two fluids undergoing mixing 
are themselves incompressible, since molecular mixing generically changes the specific volume of the mixture, the 
velocity fields of such `variable density' (VD) flows, (following the nomenclature suggested by \cite{LR2007}) 
are in general not divergence-free. This is definitely the case when the two densities are sufficiently different such 
that the Boussinesq approximation may not be applied. Commonly, the Boussinesq approximation is applied when
 the Atwood number $At$, defined as
\bel{eq:adef}
At = \frac{\rhos_2 - \rhos_1}{\rhos_2 + \rhos_1}\,,
\ee
is small\,; i.e, $At \ll 1$.
However, as discussed in detail in \cite{LR2007}, non-Boussinesq effects may occur when gradients in the density 
field become large. Following \cite{CD2001} and \cite{LR2007} the composition density $\rhos (\bx,\,t)$ of a mixture 
of two constant fluid densities $\rhos_{1}$ and $\rhos_{2}$ ($\rhos_{2} > \rhos_{1}$) is expressed in dimensionless 
form by 
\bel{sma1}
\frac{1}{\rhos(\bx,\,t)} = \frac{Y_1(\bx,\,t)}{\rhos_{1}} + \frac{Y_2(\bx,\,t)}{\rhos_{2}}\,,
\ee
where $Y_{i}(\bx,\,t)$ ($i=1,2$) are the mass fractions of the two  fluids and $Y_1+Y_2=1$. (\ref{sma1}) shows that 
the composition density $\rhos$ is bounded by 
\bel{sma2}
\rhos_{1} \leq \rhos(\bx,\,t) \leq \rhos_{2}\,.
\ee
Assuming that there is Fickian diffusion, the mass transport equations for the two species are
\bel{eq:mass}
\partial_{t}\left( \rho Y_i \right) 
+ \bnabla \cdot \left( \rhos Y_i \bu \right) = Pe_{0}\bnabla \cdot \left(\rho\bnabla Y_i \right)\,,
\ee
where $Pe_{0}$ is the P\'eclet number\,: the dimensionless Reynolds, Schmidt and P\'eclet numbers are defined in Table 
\ref{tab:param}. Since the specific volume $1/\rhos$ changes due to mixing, a non-zero divergence is induced in the velocity 
field (see Appendix A).
\bel{eq:divu}
\bnabla \cdot \bu = - Pe_{0}\Delta\left (\ln \rhos \right)
= - \frac{Pe_{0}}{\rhos} \Delta\rhos + \frac{Pe_{0}}{\rho^{*2}} |\bnabla \rhos|^{2}\,,
\ee
while summing (\ref{eq:mass}) over the two species yields the conventional continuity equation for mass conservation
\bel{eq:masscont}
\partial_{t}\rhos + \bnabla \cdot \left (\rhos \bu \right) = 0\,.
\ee
As discussed in \cite{LR2007}, the Boussinesq approximation, leading to the requirements that 
the velocity field is divergence-free $\bnabla \cdot \bu =0$ and the mass conservation equation
becomes 
\bel{eq:bouss_mass}
\partial_{t}\rhos + \bu\cdot\bnabla\rhos = Pe_{0}\Delta \rhos\,,
\ee
relies on the requirement that the second (nonlinear) term on the right hand side of (\ref{eq:divu}) can be ignored 
compared to the first term, i.e. that
\bel{eq:boussapprox}
|\bnabla \rhos|^2 \ll \rhos|\Delta\rhos|\,.
\ee
As noted by \cite{LR2007}, this condition may be violated if substantial gradients develop in the density field. It is not 
{\it a priori} clear, even when the Atwood number is very small, that  the non-divergent nature of the velocity field 
qualitatively changes the properties of the turbulent flow in ways which are significant to the mixing, and specifically 
whether regions in the flow may develop where  the condition (\ref{eq:boussapprox}) is violated. This issue can be 
explored by careful numerical simulation, as reviewed by \cite{L2013}, with a key observation (see \cite{LR2007} for 
more details) being that the pressure distribution is substantially modified by non-Boussinesq effects. 

Furthermore, the central role played by intermittency and anisotropy, as discussed in \cite{LR2008} suggests that it 
would be instructive to focus carefully on the time-dependent evolution of nonlinearity within such buoyancy-driven, 
variable density flows. Recently, a new method to assess the evolution (and depletion) of nonlinearity within turbulent 
flows has been developed centred on consideration of appropriately dimensionless $L^{2m}$ norms of the vorticity 
$\bom=\bnabla \times \bu$ and of the gradient $\bnabla\theta$ where
\bel{theta1}
\theta = \ln (\rhos/\rhos_{0})\qquad\mbox{with}\qquad
\rhos_{0} = \frac{1}{2}\left(\rhos_{1}+\rhos_{2}\right)\,.
\ee 
These $L^{2m}$-norms are scaled by an exponent ($\alpha_{m}= 2m/(4m-3)$), the origin of which comes from symmetry 
considerations for the three-dimensional Navier-Stokes equations (\cite{DGKGPV13,GDKGPV14,JDGIMA2015}). These ideas are 
explained in \S\ref{defns} and \S\ref{D1theta}.

We have been able to calculate these various scaled norms through a re-analysis of a dataset of D. Livescu, arising from 
the simulation of a buoyancy-driven flow very similar to that reported in \cite{LR2007}, which is freely available at the 
Johns Hopkins Turbulence Database (JHTDB). Using this re-analysis, there are three central questions which we wish to 
answer as the primary aims of  this paper. 
First, can the analysis approach described in \cite{DGKGPV13,GDKGPV14} be usefully generalised to consider the gradient 
of the density field, as that is naturally closely related to the buoyancy-driven mixing within the flow? Second, 
if such a generalisation can be made, can the growth of gradients in the density 
field be bounded or controlled in any meaningful way, as such bounds could yield valuable insights into the structure
and regularity of the density field and the uniform validity of the Boussinesq approximation for flows with $At \ll 1$, 
which may explain the `efficiency' of  mixing associated with buoyancy-driven turbulence? Third, does buoyancy-driven 
turbulence exhibit similar nonlinear depletion in the velocity field to the constant-density flows previously considered 
in \cite{DGKGPV13}? To address these questions,  the rest of the paper is organised as follows. In section \ref{sec:jh}, 
we describe in detail the properties of the simulation data set which we re-analyse, and we then present the results of 
this re-analysis in section \ref{sec:theta}. Finally, we draw our conclusions in section \ref{sec:conc}.

\section{Description of the database}\label{sec:jh}

As noted in the introduction, to study nonlinear depletion in buoyancy-driven turbulence we use the Johns Hopkins Turbulence Database (JHTDB) \citep{JHTDB}, a publicly available direct numerical simulation (DNS) database. For more information, please see \url{http://turbulence.pha.jhu.edu/}.

The equations used for this problem are the miscible two-fluid incompressible Navier-Stokes equations given by\,:

\beq{gov_eqn}
\partial_{t}\rhos + (\rhos u_j)_{,j} &=& 0 \\   
\partial_{t}(\rhos u_i) + (\rhos u_i u_j)_{,j} &=& -p_{,i} + \tau_{ij,j} + \frac{1}{Fr^2}\rhos g_{i} \\
u_{j,j} &=& -\frac{1}{Re_0 Sc} \left(\ln \rhos\right)_{,jj}\label{divlog}\\
\tau_{ij} &=& \rhos Re^{-1}_{0}\left(u_{i,j}+u_{j,i} - \twothirds \delta_{ij}u_{k,k}\right)\label{divergence}
\eeq
where $\rhos$ is the non-dimensional density of the mixture. 

For this problem the individual densities of the two components, $\rhos_1$ and $\rhos_2$, are constant but due to changes in mass fractions of each species, the density of the mixture can change (\ref{sma1}). For this reason, the divergence of velocity is dependent on the density as seen in equation (\ref{divlog}). The variable-density version of the petascale CFDNS code \citep{CFDNS} was used to carry out the direct numerical simulation on $1024^3$ grid points (for more information on a similar numerical study, refer to \citet{LR2007}). The Atwood number, $At$ that characterizes the density difference, is 0.05 and represents a small departure from the Boussinesq approximation. Some of the other important simulation parameters are displayed in Table \ref{tab:param}, where $U_{0}$ is the reference velocity scale, $\mu_{0}$ is the dynamic viscosity and $D$ is the mass diffusivity. 

\rem{
 The Pecl\'et number is defined by $Pe_{0} = \Rey_{0}Sc$ where the Schmidt number is defined by $Sc = \mu_{0}/D\rhos_{0}$ and the Reynolds number by $\Rey_{0} = \rho_{0}L_{0}U_{0}/\mu_{0}$\,; $L_{0}$.}

	\begin{table} \label{tab:param}
  \begin{center}
 \def~{\hphantom{0}}
  \begin{tabular}{llc}
      Reynolds number & $\Rey_0 = \rhos_{0}L_{0}U_{0}/\mu_{0}$   & 12500 \\
      Froude number & $Fr = U_{0}/\sqrt{gL_{0}}$   & 1 \\
      Schmidt  number & $Sc =  \mu_{0}/D\rhos_{0}$  & 1 \\
      Pecl\'et number & $Pe_{0} = \Rey_{0}Sc$   & 12500\\
      Atwood number & $At = (\rhos_2-\rhos_1)/(\rhos_2+\rhos_1)$   & 0.05 \\
			Domain length & $L$ & $2\pi$\\
			Non-dimensionalization length & $L_{0}$ & $1$
  \end{tabular}
  	\caption{Simulation parameters}
  \end{center}
\end{table}

In the beginning, the fluids are initialized as random blobs with periodic boundary in each direction and an initial diffusion layer at the interface. At sufficiently late time, the statistically homogeneous turbulent flow generated by such conditions resembles the interior of the mixing layer (away from the wall and/or edge effects) of the Rayleigh-Taylor instability at the turbulent stage \citet{LR2007}. 

The inhomogeneities in the transport terms are important at the edge and thus, it is safe to assume that the homogeneous simulation data under consideration describes the core of a fully developed mixing layer. Eventually, the turbulent behaviour dies out as the fluids become mixed at the molecular level. 

This high resolution data is stored as a sequence of 1011 files each representing $32^3$ spatial points at each time step starting from $t = 0$ to $t = 40.44$. The velocity gradients in the database are calculated as a post-processing step using a 4th order central finite differencing approximation from the data. 

If the gradients or the state variables are desired at a particular spatial location between the stored grid points, 4th order spatial interpolation or the 6th order Lagrangian interpolation are used. To get the temporal values other than the stored ones, a piecewise cubic harmonic interpolation is employed.

\section{Results}\label{sec:theta}

\subsection{Definitions}\label{defns}

It is clear from (\ref{sma1}) that the composition density $\rhos$ is bounded by $\rhos_{1} \leq \rhos \leq \rhos_{2}$. 
Moreover, in Appendix \S\ref{appA} it is also shown that $\|\rhos\|_{L^{\infty}}$ is bounded above by its initial data 
provided the advecting $\bu$-field is regular. However, our interest lies more in $\bnabla\rhos$, but it is difficult to 
work with this quantity alone. To circumvent this problem, it is shown in Appendix \S\ref{appB} that with a normalization 
density $\rhos_{0} = \shalf(\rhos_{1} +\rhos_{2})$, the new variable $\theta$ defined by
\bel{thetadef}
\theta(\bx,\,t) = \ln \rho(\bx,\,t)\qquad\qquad \rho = \frac{\rhos}{\rhos_{0}}\,,
\ee
changes the evolution equation for $\rhos$ into a deceptively innocent-looking diffusion-like equation 
\bel{dep1a}
\left(\partial_{t} + \bu\cdot\bnabla\right)\theta = Pe_{0}^{-1}\Delta\theta\,,
\ee
but with an equation for $\bnabla \cdot \bu$ that depends on two derivatives of $\theta$
\bel{dep1b}
\bnabla\cdot\bu = - Pe_{0}^{-1}\Delta\theta\,.
\ee
It is now easier to work with $\theta = \ln\rho$ evolving according to (\ref{dep1a}) and (\ref{dep1b}) 
by considering both $\bnabla\theta$ and $\bom = \mbox{curl}\,\bu$ in the higher norms $L^{2m}
\left(\mathcal{V}\right)$ defined by ($1 \leq m < \infty$)
\beq{eq:omo}
\Omega_{m,\theta} &=& \left(\left(L/L_{0}\right)^{3}\I |\bnabla\theta|^{2m}dV\right)^{1/2m}\,,\\
\Omega_{m,\omega} &=& \left(\left(L/L_{0}\right)^{3}\I|\bom|^{2m}dV\right)^{1/2m}\,,\label{eq:omt}
\eeq
where $L_{0}$ is the non-dimensionalization length in the  JHT-database. The natural sequence of H\"older 
inequalities 
\bel{Omin1}
\Omega_{m,\theta} \leq (L/L_{0})^{3/2m(m+1)}\Omega_{m+1,\theta}\,,
\ee
has a multiplicative factor which is only unity when $L=L_{0}$.  If we define
\bel{alphadef}
\alpha_{m} = \frac{2m}{4m-3}\,
\ee 
then the exponent on $L/L_{0}$ in (\ref{Omin1}) is related to $\alpha_{m}$ and $\alpha_{m+1}$ by
\bel{alpha1}
\frac{3}{2m(m+1)} = \frac{1}{\alpha_{m+1}} - \frac{1}{\alpha_{m}}\,.
\ee
In turn, this leads us to define a natural dimensionless length
\bel{defnsmall}
\ell_{m} =  \left(L/L_{0}\right)^{1/\alpha_{m}}\,,
\ee
which turns (\ref{Omin1}) into $\ell_{m}\Omega_{m,\theta} \leq \ell_{m+1}\Omega_{m+1,\theta}$.
The aim is to assume there exists a solution of (\ref{dep1a}) in tandem with the vorticity 
field $\bom$. Motivated by the depletion properties studied in Donzis \textit{et al} (2013) 
and Gibbon \textit{et al} (2014) for the Navier-Stokes equations, the following definitions 
are made
\bel{dep2a}
D_{m,\theta} = \left(\ell_{m}\Omega_{m,\theta}\right)^{\,\alpha_{m}}\,,
\ee
\bel{dep2b}
D_{m,\omega} = \left(\ell_{m}\Omega_{m,\omega}\right)^{\,\alpha_{m}}\,.
\ee
The $\alpha_{m}$-scaling in (\ref{dep2a}) and (\ref{dep2b}) has its origins in scaling properties of the 
three-dimensional Navier-Stokes equations (see \cite{GDKGPV14}). 
Note that the ordering observed in (\ref{Omin1}) does not necessarily hold for the $D_{m,\theta}$ 
or the $D_{m,\omega}$ because $\alpha_{m}$ \textit{decreases} with $m$. In the JHT-database 
the dimensionless domain size is $2\pi$ thus indicating that $L/L_{0} = 2\pi$.

\subsection{The evolution of $D_{1,\theta}$}\label{D1theta}

\begin{figure}
\begin{center}
\includegraphics[width=0.32\columnwidth]{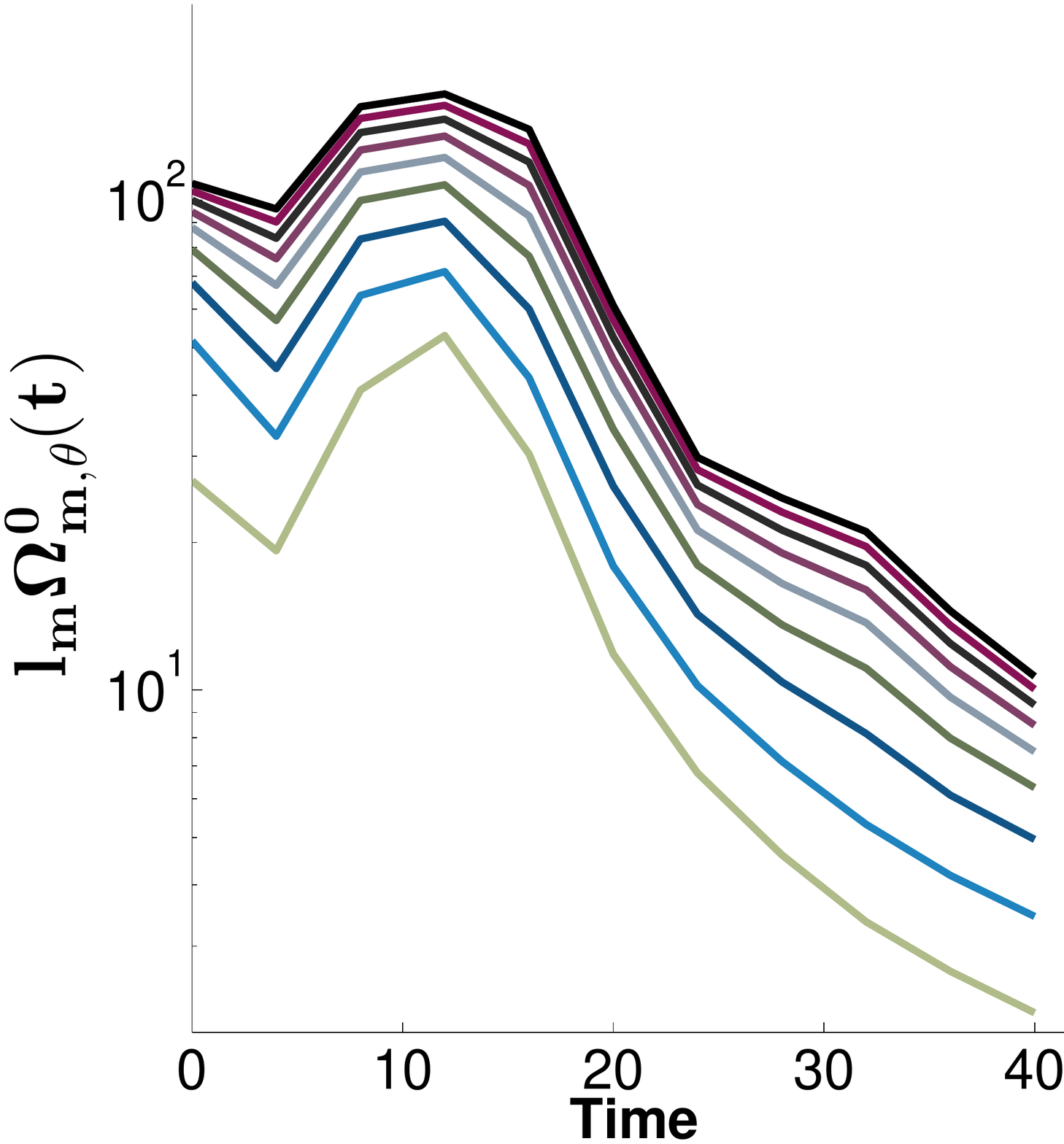}
\includegraphics[width=0.32\columnwidth]{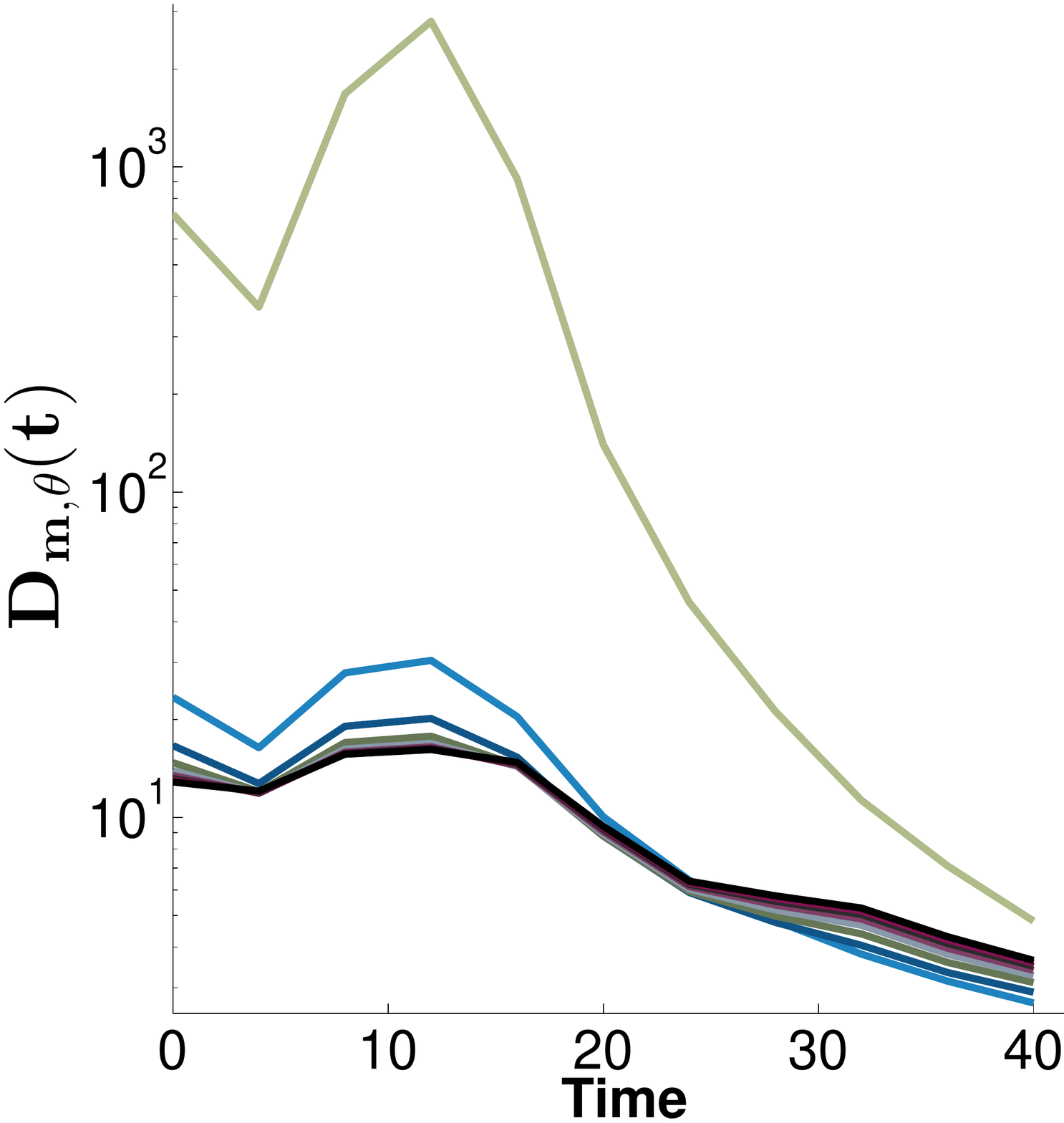}
\includegraphics[width=0.32\columnwidth]{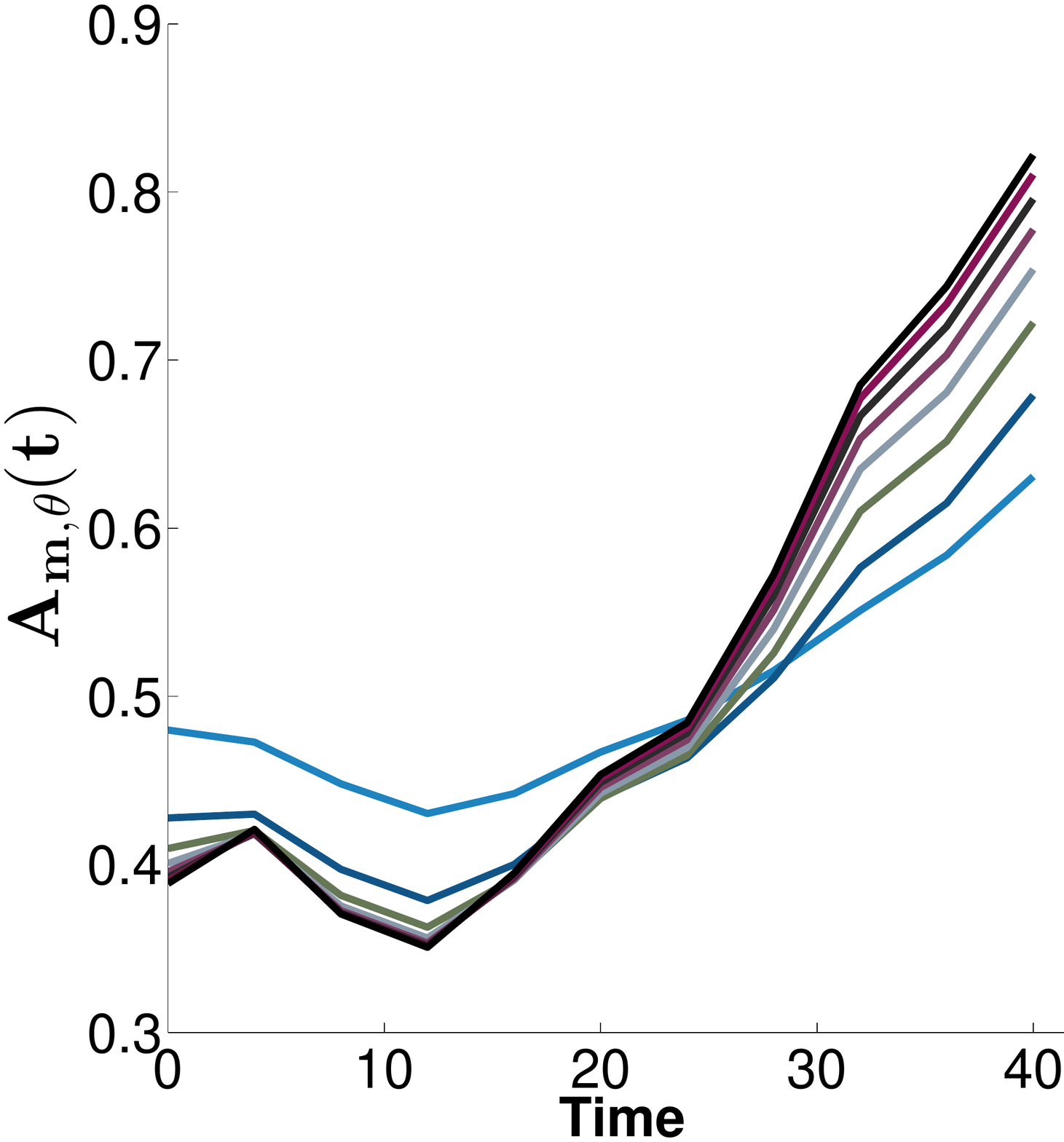}\\
\includegraphics[width=0.70\columnwidth]{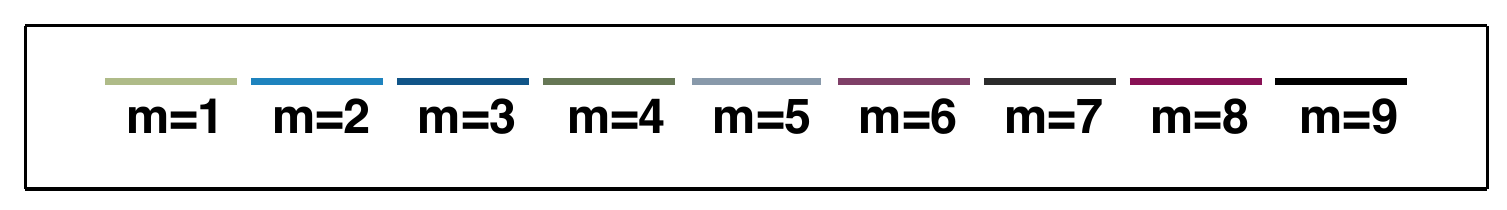}
\end{center}
\caption{Time variation of: (\textit{a}) $l_m\Omega_{m,\theta}(t)$, as defined in (\ref{eq:omt})\,; 
(\textit{b}) $D_{m,\theta}(t)$, as defined in (\ref{dep2a})\,; (\textit{c}) $A_{m,\theta}(t)$ as defined 
in (\ref{Amdef2}). }
\label{fig:fig1}
\end{figure}

\begin{figure}
\begin{center}
\begin{tabular}{cc}
\includegraphics[width=0.48\hsize]{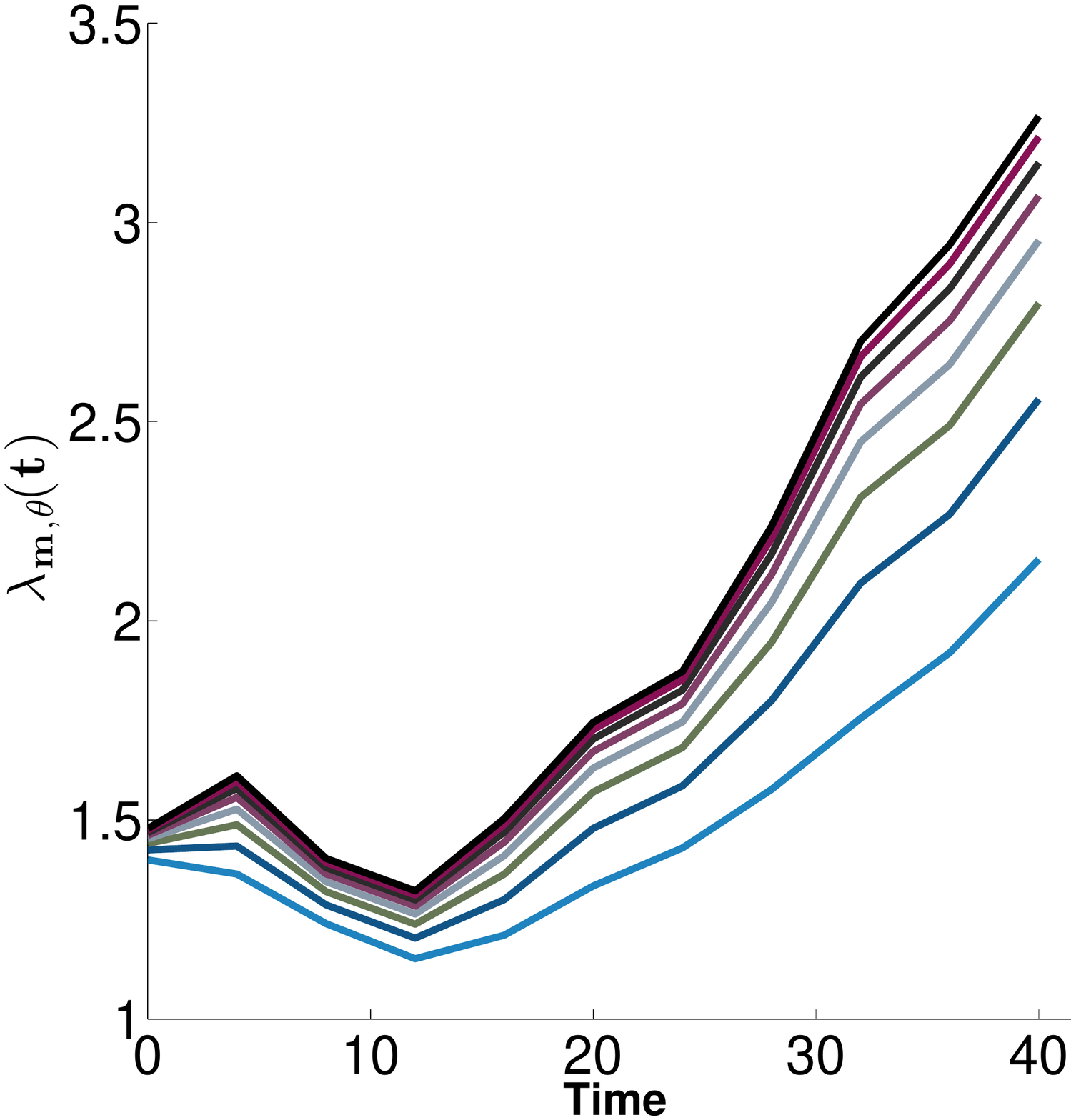}
\includegraphics[width=0.48\hsize]{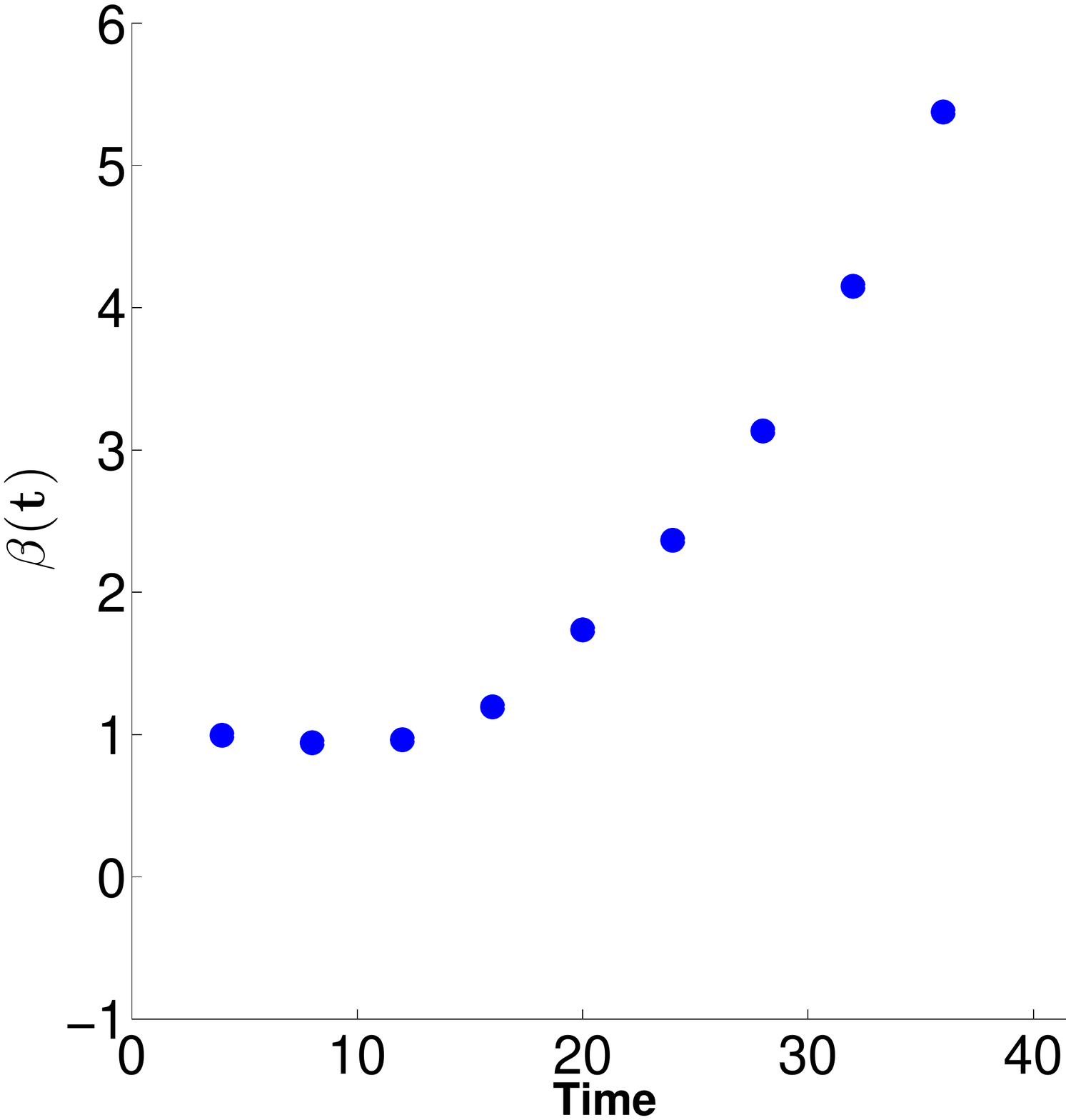}
\end{tabular}
\end{center}
\caption{Time variation of: (\textit{a}) $\lambda_{m,\theta}(t)$, as defined in (\ref{lamdef}), which fan out and grow with time\,; 
(\textit{b}) $\beta(t)$ as defined in (\ref{dep11}).}
\label{fig:fig2}
\end{figure}

Now formally consider the time evolution of $D_{1}$ using (\ref{dep1a})
\bel{dep3a}
\frac{1}{2}\frac{d~}{dt}\I |\bnabla\theta|^{2}dV = 
\I \I \bnabla\theta\cdot\left(Pe_{0}^{-1}\Delta - \bnabla\bu \right)\cdot\bnabla\theta\,dV
+ \frac{1}{2} \I |\bnabla\theta|^{2}(\bnabla\cdot\bu)\,dV\\
\ee
and so, integrating by parts and using (\ref{dep1b}), we have
\bel{dep3b}
\frac{1}{2}\frac{d~}{dt}\I |\bnabla\theta|^{2}dV \leq  
- Pe_{0}^{-1}\I |\Delta\theta|^{2}dV  + \I |\bnabla\theta|^{2}|\nabla\bu|\,dV 
+ \frac{1}{2} Pe_{0}^{-1}\I |\bnabla\theta|^{2}|\Delta\theta|\,dV\,.
\ee
For $m \geq 2$, and noting that $\frac{m-2}{2(m-1)} + \frac{m}{2(m-1)} =2$, consider the term 
\beq{dep4}
\I |\bnabla\theta|^{2}|\bnabla\bu|\,dV  
&\leq&  \Omega_{1,\theta}^{\frac{m-2}{m-1}}\Omega_{m,\theta}^{\frac{m}{m-1}}\Omega_{1,\omega}\nonumber\\
&=& c_{1,m}\,D_{1,\theta}^{\frac{m-2}{2(m-1)}}D_{m,\theta}^{\frac{m}{\alpha_{m}(m-1)}}D_{1,\omega}^{1/2}\,.
\eeq
where the factors of $\ell_{m}$ and $2\pi$ have been absorbed into the
dimensionless constant $c_{1,m}$. Now we turn 
to an idea introduced for the three-dimensional Navier-Stokes equations by \cite{GDKGPV14} in which it was discovered 
that a relation between $D_{m}$ and $D_{1}$ fitted the data. In \cite{GDKGPV14} the formulae in (\ref{Amdef2}) and 
(\ref{lamdef}) were found to fit the maxima in time of the $D_{m}$ versus $D_{1}$ curves with $\lambda$ approximately 
constant. However, in a subsequent paper \cite{JDGIMA2015} it has been shown that these formulae have a rigorous basis 
if the set of exponents $\{\lambda_{m}(t)\}$ are allowed to be time dependent. Following this, the JHT-database shows that the 
relation between $D_{m}$ and $D_{1}$ takes the form 
\bel{Amdef2}
D_{m,\theta}(t) = D_{1,\theta}^{A_{m,\theta}(t)}\,
\ee
The data is consistent with $A_{m,\theta}(t)$ being expressed as
\bel{lamdef}
 = \frac{\ln D_{m,\theta}}{\ln D_{1,\theta}} \equiv A_{m,\theta}(t) = \frac{\lambda_{m,\theta}(t)(m-1) + 1}{4m-3}\,,
\ee
Plots of $\ell_{m}\Omega_{m,\theta}(t)$, $D_{m,\theta}(t)$ and $A_{m,\theta}$ are shown in figure \ref{fig:fig1}, with plots 
of the corresponding $\lambda_{m,\theta}(t)$ in figure \ref{fig:fig2}a.  Note that the set $\{\lambda_{m,\theta}(t)\}$ fan out 
with time with no tendency to coincide. Nonlinear depletion occurs
when $A_{m,\theta} < 1$, which figure 1 shows is the case. 
\par
\begin{figure}
\begin{center}
\begin{tabular}{c}
\includegraphics[width=0.5\hsize]{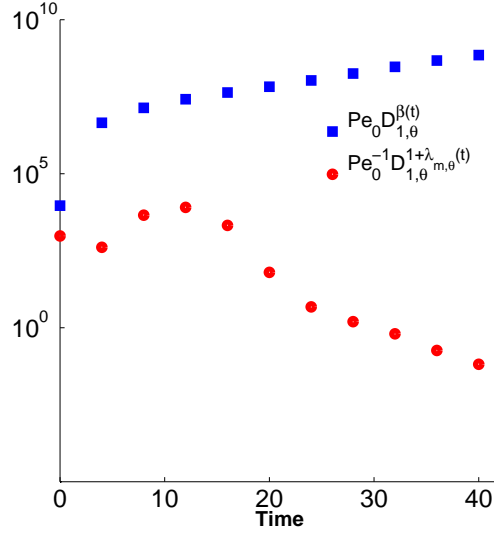}
\end{tabular}
\end{center}
\caption{
Time variation of $Pe_{0}D_{1,\theta}^{\beta(t)}$ (plotted with blue squares) and 
$Pe_{0}^{-1}D_{1,\theta}^{1+\lambda_{m,\theta}(t)}$-term (plotted with red circles) where 
$\lambda_{m,\theta}(t)$ is chosen to be the maximum value over $m$ at each time step with $Pe_0 = 12,500$.}
\label{fig:fig3}
\end{figure}

\smallskip
Inserting (\ref{Amdef2})  into the right hand side of (\ref{dep4}) gives
\beq{dep5}
\I |\bnabla\theta|^{2}|\bnabla\bu|\,dV  &\leq& c_{1,m}D_{1,\omega}^{1/2}D_{1,\theta}^{(1+\lambda_{m,\theta})/2}\nonumber\\
&\leq &\frac{1}{2} Pe_{0} D_{1,\omega} + c_{2,m} Pe_{0}^{-1}D_{1,\theta}^{1+\lambda_{m,\theta}}\,.
\eeq
where the use of a H\"older inequality has split up the terms on the last line of the right hand side.
The same idea is used on the last term in (\ref{dep3b}) with $|\bnabla\bu|$ replaced by $|\Delta\theta|$\,:
\beq{dep6}
Pe_{0}^{-1}\I |\bnabla\theta|^{2}|\Delta\theta|\,dV &\leq& \left(Pe_{0}^{-1}\|\Delta\theta\|_{2}^{2}\right)^{1/2}
\left(2c_{3,m}Pe_{0}^{-1} D_{1,\theta}^{1+\lambda_{m,\theta}}\right)^{1/2}\nonumber\\
&\leq& \frac{1}{2} Pe_{0}^{-1}\|\Delta\theta\|_{2}^{2} + c_{3,m}Pe_{0}^{-1} D_{1,\theta}^{1+\lambda_{m,\theta}}\,.
\eeq
Altogether, (\ref{dep3b}) becomes
\bel{dep7}
\frac{1}{2}\dot{D}_{1} \leq -\frac{1}{2} Pe_{0}^{-1}\|\Delta\theta\|_{2}^{2} 
+ c_{4,m} Pe_{0}^{-1} D_{1,\theta}^{1+\lambda_{m,\theta}} + \frac{1}{2} Pe_{0}D_{1,\omega}\,.
\ee
A simple integration by parts shows that 
\bel{dep8}
\|\bnabla\theta\|_{2}^{2} \leq \|\Delta\theta\|_{2}\|\theta\|_{2}
\ee
and so we have 
\bel{dep10}
\frac{1}{2}\dot{D}_{1,\theta} \leq -\frac{1}{2} Pe_{0}^{-1}\frac{D_{1,\theta}^{2}}{\|\theta\|_{2}^{2}}
+ c_{4,m} Pe_{0}^{-1} D_{1,\theta}^{1+\lambda_{m,\theta}(t)}+ \frac{1}{2} Pe_{0} D_{1,\omega}\,.
\ee
Because $\rhos$  is bounded both below and above then so is $\|\theta\|_{2}^{2}$. Thus the competition on the right 
hand side of (\ref{dep10}) in powers of $D_{1,\theta}$ lies between the negative $D_{1,\theta}^{2}$ term and either 
$Pe^{-1}D_{1,\theta}^{1+\lambda_{m,\theta}}$ or the $Pe_{0}D_{1,\omega}$ terms. To turn the differential inequality 
(\ref{dep10}) into one in $D_{1,\theta}$ alone requires a relation between and $D_{1,\theta}$ and $D_{1,\omega}$, 
with the latter representing the fluid vorticity. Analytically, we have been unable to establish a relation between them 
but the JHT database provides us with the relation 
\bel{dep11}
D_{1,\omega} = D_{1,\theta}^{\beta(t)}\,,
\ee
where the growth in the exponent $\beta(t)$ is shown in figure 2b. Moreover, figure \ref{fig:fig3} shows that the 
$Pe_{0}D_{1,\theta}^{\beta(t)}$-term (plotted with blue squares) in (\ref{dep7}) is dominant over the 
$Pe_{0}^{-1}D_{1,\theta}^{1+\lambda_{m,\theta}(t)}$-term (plotted with red circles), even when $\lambda_{m,\theta}(t)$ 
is chosen to be the maximum across $m$ at  each particular time step. The plots of  $1+\lambda_{m,\theta}$ 
and $\beta(t)$ both show that the values of these two quantities are both greater than $2$ and thus cannot be controlled 
by the $-D_{1,\theta}^2$ term in (\ref{dep10}). $D_{1,\theta}$ is bounded only for extremely short times. Thus the possibility 
of the blow-up of $D_{1,\theta}$ in a finite time cannot be discounted.
\par
\begin{figure}
\begin{center}
\begin{tabular}{ccc}
\includegraphics[width=0.32\columnwidth]{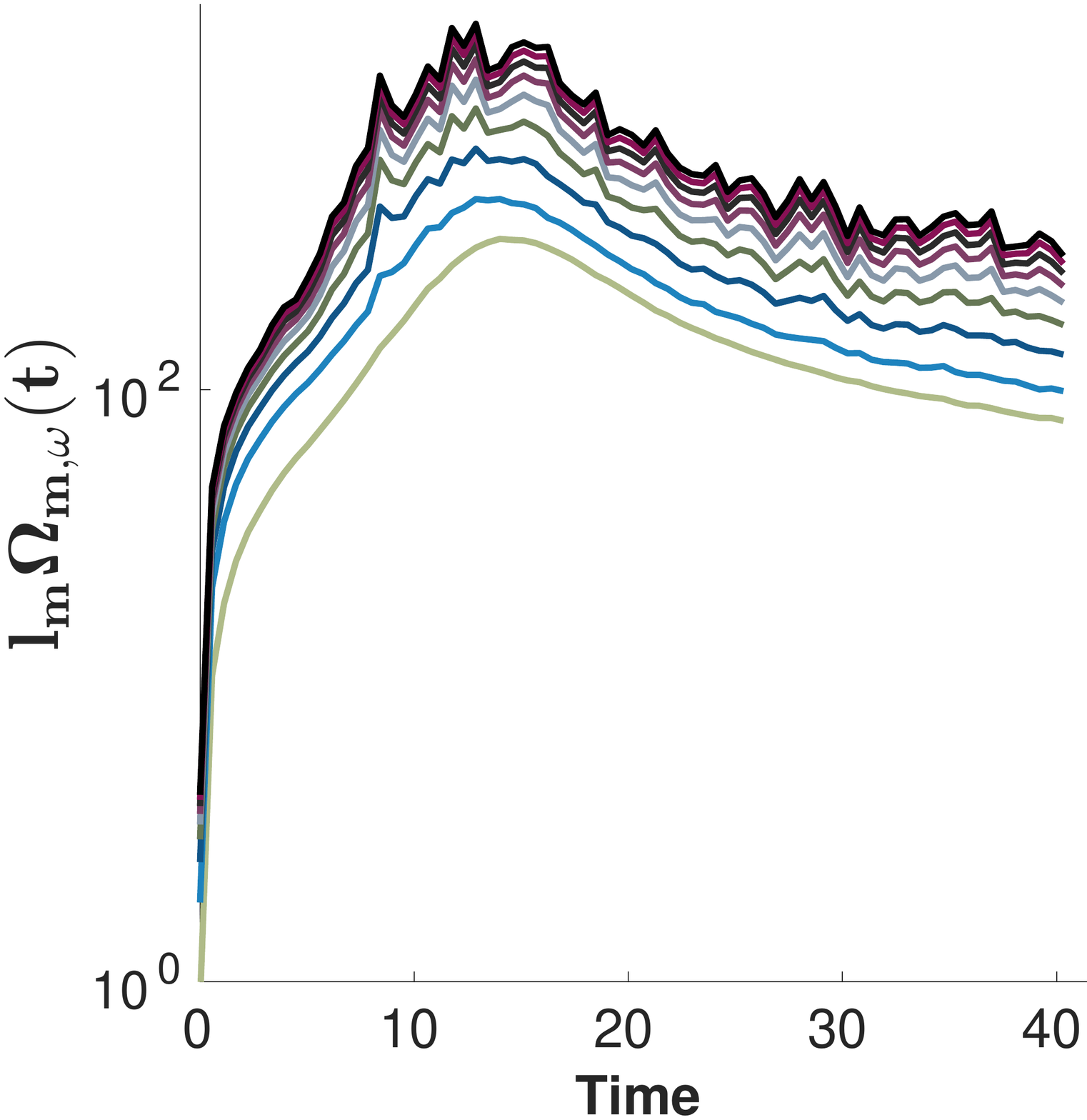}
\includegraphics[width=0.32\columnwidth]{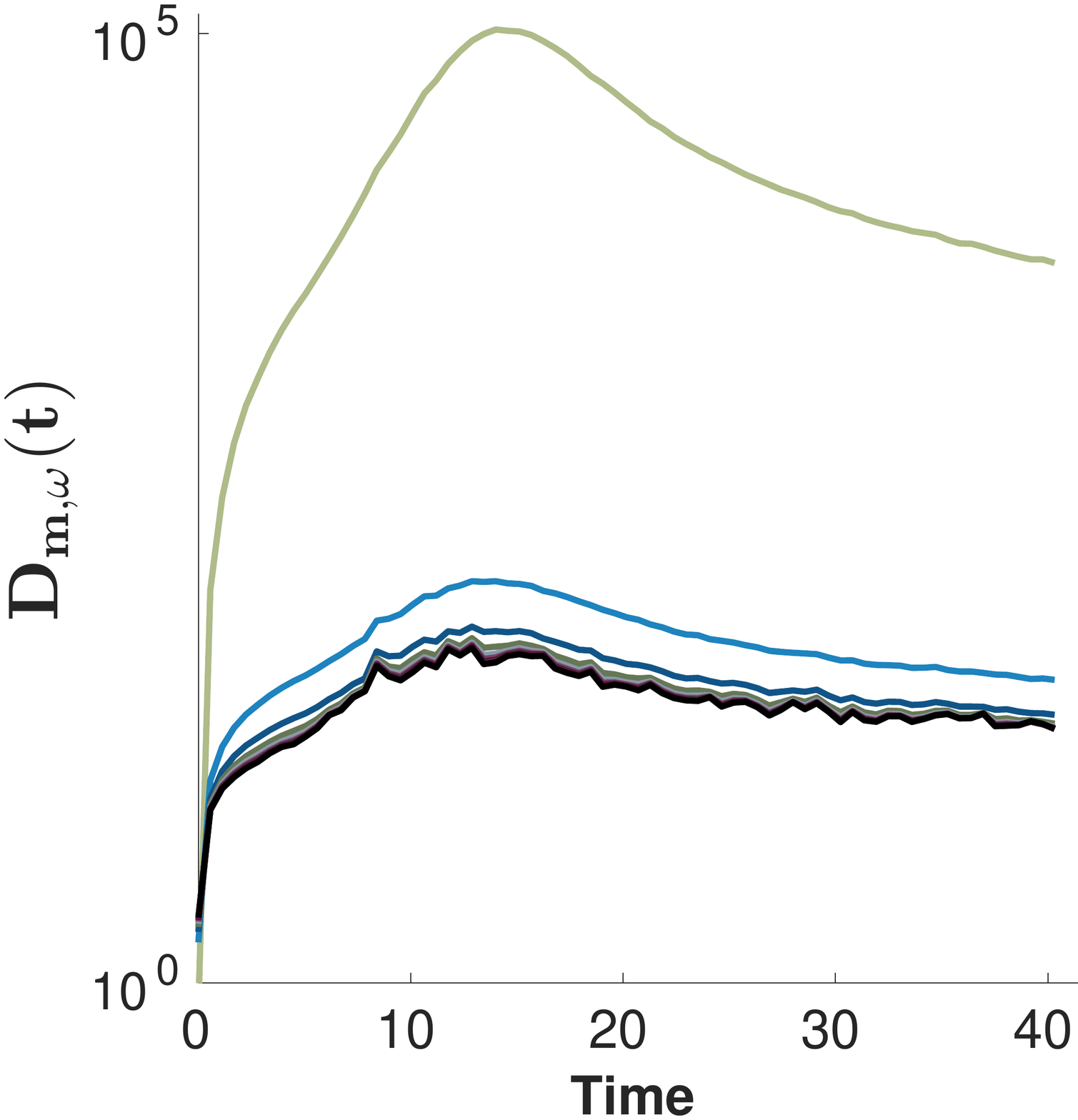}
\includegraphics[width=0.32\columnwidth]{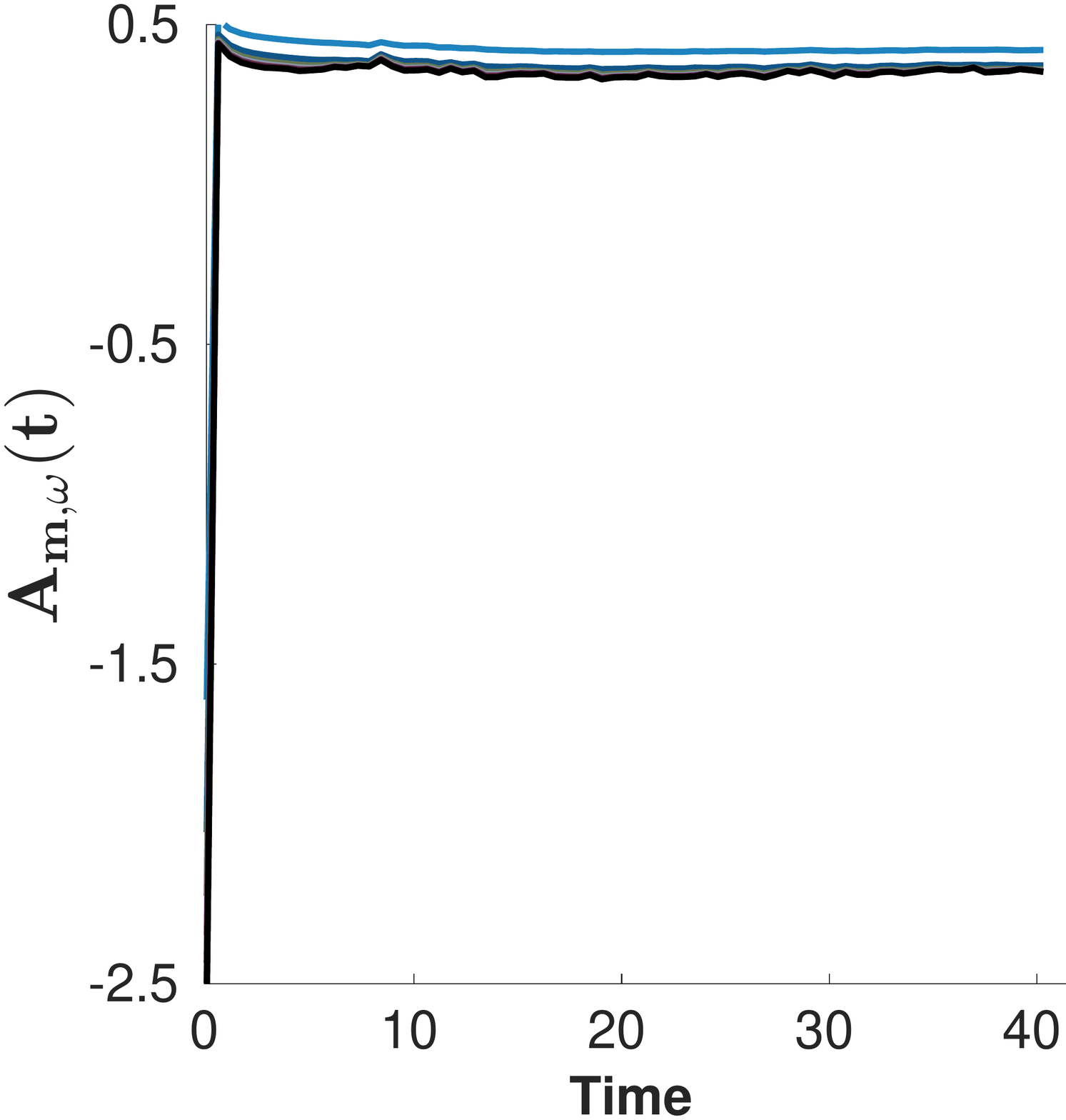}\\\\
\includegraphics[width=0.7\columnwidth]{color_bar.pdf}
\end{tabular}
\end{center}
\caption{Time variation of: (\textit{a}) $l_m\Omega_{m,\omega}(t)$ as defined in (\ref{eq:omt})\,; (\textit{b})
$D_{m,\omega}(t)$, as defined in (\ref{dep2b})\,; (\textit{c}) $A_{m,\omega}(t)$, as defined in (\ref{eq:lamodef}).}
\label{fig:fig4}
\end{figure}

\begin{figure}
\begin{center}
\begin{tabular}{cc}
\includegraphics[width=0.45\hsize]{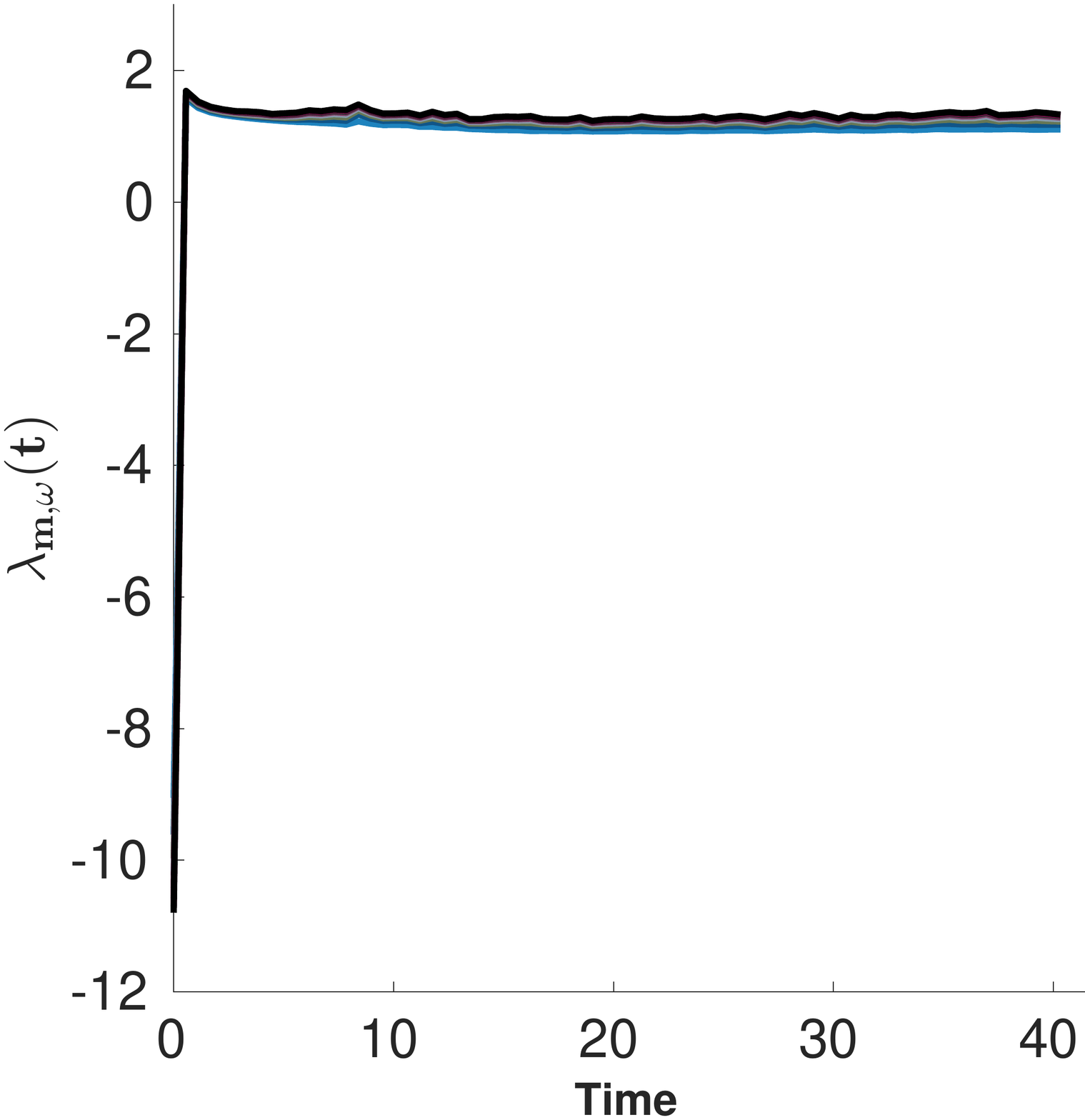}
\end{tabular}
\end{center}
\caption{Time variation of $\lambda_{m,\omega}(t)$, calculated using the relation 
(\ref{eq:lamodef}).}
\label{fig:fig5}
\end{figure}

\smallskip
Finally, figure \ref{fig:fig4} shows the equivalent set of plots of the time variation of $\ell_{m}\Omega_{m,\omega}(t)$, 
(as defined in (\ref{eq:omt}) and (\ref{dep2b})), $D_{m,\omega}(t)$ (as defined in (\ref{dep2b})) and $A_{m,\omega}(t)$
defined as 
\bel{eq:amodef}
A_{m,\omega}(t) = \ln D_{m,\omega}/\ln D_{1,\omega}.
\ee
In figure \ref{fig:fig5}, we  also show the time variation of the  corresponding $\lambda_{m,\omega}(t)$,
calculated using the analogous relationship
\bel{eq:lamodef}
A_{m,\omega}(t) = \frac{\left[\lambda_{m,\omega}(t) (m-1) + 1\right]}{(4m-3)}\,.
\ee
It is apparent that the turbulent fluid part of the problem, which drives and dominates the system, has 
corresponding $\lambda_{m,\omega}(t)$ that are flat in time and sit in the range $1 < \lambda_{m,\omega} 
< 2$. This is consistent with the behaviour found in three-dimensional Navier-Stokes flow described in Donzis \textit{et al} 
(2013), Gibbon \textit{et al} (2014) and Gibbon (2015). Note that this contrasts strongly with the behaviour of 
the $\theta$-variable where the $\lambda_{m,\theta}$ fan out and grow in time, as shown in figure \ref{fig:fig2}.

\section{Conclusion}\label{sec:conc}

The numerical evidence in figure 2a suggests strong growth in $\lambda_{m,\theta}(t)$ which is consistent with 
strong growth in $\bnabla\rhos$ even while $\rhos$ is bounded. There are varying degrees of 
nonlinear depletion in the sense that $A_{m,\theta} < 1$, and $A_{m,\omega} < 1$ (as in figure 4c and 5). 
Depletion in $A_{m,\theta}$  reduces as the growth of $\lambda_{m,\theta}$ to the value $3.5$ in the final stages attests. 
Indeed, note that $\lambda_{m,\theta} =  4$ would give a linear relation and be equivalent to a full estimate of the 
nonlinearity. Depletion in $D_{1,\omega}$ is quite severe, as shown in figures 4c and 5, which is consistent with the 
same effect observed in Navier-Stokes flows. Despite this, the cross-effect of the turbulent fluid flow driving the growth 
of $D_{1,\theta}$ through the exponent $\beta(t)$ swamps the term $D_{1,\theta}^{1+\lambda_{m,\theta}}$ 
in (\ref{dep10}). 
\par\smallskip
Following \cite{LR2007}, there is another way of looking at the growth in $\bnabla\rhos$. Consider the equation 
for $\theta$ and introduce a new velocity field $\bv = \bu + Pe_{0}\bnabla\theta$. The Hopf-Cole-like 
transformation $\theta = \ln\rho$ in (\ref{thetadef}) then leads to an exact cancellation of the nonlinear terms 
in (\ref{dep1a}) to give
\bel{cm8}
\left(\partial_{t} + \bv\cdot\bnabla\right)\rho = Pe_{0}^{-1}\Delta\rho \,,
\qquad\mbox{with}\qquad  \bnabla . \bv = 0\,.
\ee
This is the linear advection diffusion equation driven by a divergence-free velocity field. Note that $\bom 
= \mbox{curl}\,\bu =  \mbox{curl}\, \bv$. The fact that $\bv$ is
actually an (explicit) function of $\bnabla\theta$ makes (\ref{cm8}) less 
simple than it first appears. 

Nevertheless, this equation provides a
hint as to how we might look at the dynamics in a 
descriptive way. 
Consider a one-dimensional horizontal section through a rightward moving wave of $\rhos$ at a snapshot 
in time\,: in the frame of the advecting velocity $\bu$ the relevant component of $\bv$ is greater on 
the back face of any part of the wave (where $\bnabla\rhos > 0$) than on the front face (where $\bnabla
\rhos < 0$). Thus in the advecting frame, (\ref{cm8}) implies that not only is there the usual advection and 
diffusion but also a natural tendency for the back of a wave to catch up with the front, thus leading to 
steepening of $\bnabla\rhos$. 
This is consistent with the evidence from (\ref{dep10}) which 
leaves open the possibility that $D_{1,\theta}$ could blow up after a finite time or at least grow sufficiently 
strongly that the mixing is driven down to near molecular scales where
the validity of the model fails. Interestingly, this then hints
that buoyancy-driven turbulence may well be more
intense in some sense than constant-density turbulence, 
which may explain the observed extremely efficient 
mixing possible in such flows.

\section*{Acknowledgements}

We acknowledge, with thanks, the staff of IPAM UCLA where this collaboration began in the Autumn of 
2014 on the programme ``Mathematics of Turbulence''. We would also
like to thank C. Doering and  D. Livescu for useful discussions.
Research activity of C.P.C. is supported by    EPSRC Programme  Grant
EP/K034529/1 (``Mathematical Underpinnings of Stratified
Turbulence''). All the numerical data used is freely available from
the Johns Hopkins Turbulence Database (JHTDB) \cite{JHTDB}, a publicly available direct numerical 
simulation (DNS) database. For more information, please see \url{http://turbulence.pha.jhu.edu/}.

\appendix

\section{The equations for the composite density}\label{appB}

Following \cite{CD2001} and \cite{LR2007} the composition density $\rhos (\bx,\,t)$ of a mixture of two 
constant fluid densities $\rhos_{1}$ and $\rhos_{2}$ ($\rhos_{2} > \rhos_{1}$) is expressed by (\ref{sma1}) 
where $Y(\bx,\,t) = Y_{2}$ is the mass fraction of the heavier fluid. It is important to stress that the two fluids 
are assumed to be incompressible, yet we do not make the Boussinesq approximation and so the difference 
between the two densities is allowed to take arbitrary values. Under the transport of a (dimensionless) 
velocity field $\bu(\bx,\,t)$, $\rhos$ obeys the equation of conservation of mass
\bel{cm1a}
\partial_{t}\rhos + \bnabla \cdot  (\rhos\bu) = 0\,,
\ee
and the species transport equation
\bel{cm1b}
\partial_{t}(\rhos Y) + \bnabla \cdot \left(\rhos Y\bu + \bj_{F}\right) = 0\,,
\ee
where the divergence of the flux $\bj_{F}$ represents Fickian diffusion, i.e.
\bel{}
\bj_{F} =  - Pe_{0}^{-1}\rhos\bnabla Y\,.
\ee
where the P\'eclet number has been defined in table \ref{tab:param}. Given that the solution of (\ref{sma1}) shows 
that $\rhos Y$ is linear in $\rhos$ such that
\bel{cm2}
\rhos Y = a \rhos + b\,,\qquad a= \frac{\rhos_{2}}{\rhos_{2}- \rhos_{1}}\,,
\qquad b = - \frac{\rhos_{1}\rhos_{2}}{\rhos_{2} - \rhos_{1}}\,,
\ee
equation (\ref{cm1b}) simplifies to
\bel{cm3}
b\, \bnabla \cdot \bu = Pe_{0}^{-1}\bnabla \cdot (\rhos\bnabla Y)\,.
\ee
Noting from (\ref{cm2}) that $\rhos\bnabla Y = - b\bnabla(\ln\rhos)$ the 
coefficient $b$ cancels to make (\ref{cm3}) and (\ref{cm1a}) into\,:
\bel{cm4}
\bnabla\cdot\bu = - Pe_{0}^{-1}\Delta(\ln\rhos)\,.
\ee
\bel{cm5}
\left(\partial_{t} + \bu\cdot\bnabla\right)\rhos =  Pe_{0}^{-1}\rhos\Delta(\ln\rhos)\,.
\ee
An interesting observation is that using a normalization density $\rhos_{0}$ and with the definition
\bel{thetadefappB}
\theta(\bx,\,t) = \ln \rho \qquad\qquad \rho = \frac{\rhos}{\rhos_{0}}\,.
\ee
(\ref{cm5}) becomes a deceptively innocent-looking diffusion-like equation 
\bel{dep1appB}
\left(\partial_{t} + \bu\cdot\bnabla\right)\theta = Pe_{0}^{-1}\Delta\theta\,,
\ee
with an equation for $\bnabla \cdot \bu$ that depends on two derivatives of $\theta$
\bel{dep2appB}
\bnabla\cdot\bu = - Pe_{0}^{-1}\Delta\theta\,.
\ee
\rem{
(\ref{dep1b}) points us to the definition of a new divergence-free velocity field 
\bel{cm6}
\bv = \bu + Pe_{0}^{-1}\bnabla\theta\qquad\mbox{where}\qquad \bnabla
\cdot \bv = 0\,,
\ee
which turns (\ref{dep1a}) into
\bel{cm7}
\left(\partial_{t} + \bv\cdot\bnabla\right)\theta = Pe_{0}^{-1}\left(\Delta\theta + |\bnabla\theta|^{2}\right)\,.
\ee
The Hopf-Cole-like transformation $\theta = \ln\rho$ in (\ref{thetadef}) linearizes (\ref{cm7}) to 
\bel{cm8}
\left(\partial_{t} + \bv\cdot\bnabla\right)\rho = Pe_{0}^{-1}\Delta\rho\,,\qquad\mbox{with}\qquad  \mbox{div}\,\bv = 0\,.
\ee
This is the \textbf{linear} advection diffusion equation driven by a divergence-free velocity field. Note that $\bom = \mbox{curl}\,\bu 
=  \mbox{curl}\, \bv$. There is a twist, however, because the advecting velocity $\bv$ depends upon $\bnabla\rho/\rho$ so it is difficult 
to extract any regularity information except to say 
that (\ref{cm8})  has regular solutions provided the advecting velocity field $\bv$ is regular. To be more specific, a maximum principle 
tells us that $\rho$ is bounded provided $\bnabla\rho$ and $\bu$ are finite, which is consistent with the model.
}

\section{Proof of the boundedness of $\|\rhos\|_{L^{\infty}(\mathcal{V})}$}\label{appA}

To prove the boundedness of $\|\rhos\|_{L^{\infty}(\mathcal{V})}$ under a sufficiently regular 
advecting field $\bu$ we write
\bel{Ap1}
\frac{1}{2m}\frac{d~}{dt}\I |\rhos|^{2m}dV = - \I \rho^{*(2m-1)}\,\bnabla\cdot(\rhos\bu)\,dV\,,
\ee
and
\bel{Ap2}
\rho^{*(2m-1)}\,\bnabla\cdot(\rhos\bu) = \left(1-\frac{1}{2m}\right)\rho^{*2m}\,\bnabla\cdot\bu
+ \frac{1}{2m}\bnabla\cdot(\rho^{*2m}\bu)\,.
\ee
(\ref{Ap1}) then becomes
\bel{Ap3}
\frac{1}{2m}\frac{d~}{dt}\I |\rhos|^{2m}dV = -\left(1-\frac{1}{2m}\right)\I\rho^{*2m}\,\bnabla\cdot\bu\,dV
\ee
where the volume integral of the second term in (\ref{Ap2}) is zero through the Divergence Theorem. Using 
(\ref{divlog}), (\ref{Ap3}), becomes
\beq{Ap4}
\frac{1}{2m}\frac{d~}{dt}\I |\rhos|^{2m}dV &=& Pe_{0}^{-1}\left(1-\frac{1}{2m}\right)
\I\rho^{*2m}\,\Delta(\ln\rhos)\,dV \nonumber\\
&=& - Pe_{0}^{-1}(2m-1)\I \rho^{*2(m-1)}|\bnabla\rhos|^{2}\,dV\nonumber\\
&=&  - Pe_{0}^{-1}\frac{(2m-1)}{m^{2}}\I |\bnabla\rho^{*m}|^{2}\,dV
\eeq
so Poincar\'e's inequality shows each norm $\|\rhos\|_{L^{2m}(\mathcal{V})}$ decays exponentially from its initial
conditions. In the limit $m\to\infty$, $\|\rhos\|_{L^{\infty}(\mathcal{V})}$ is bounded by its initial conditions.
\hfill $\blacksquare$

\bibliographystyle{jfm}
\bibliography{refs}

\end{document}